\documentclass[a4paper]{PoS}
\usepackage{units}
\usepackage{amsmath, bm}
\usepackage{mathtools}
\usepackage{graphics} 
\usepackage{adjustbox}
\usepackage{multirow}
\usepackage{tabularx}

\makeatletter
\let\jnl@style=\rmfamily 
\def\ref@jnl#1{{\jnl@style#1}}%
\newcommand\aj{\ref@jnl{AJ}}
\newcommand\araa{\ref@jnl{ARA\&A}}
\newcommand\apj{\ref@jnl{ApJ}}
\newcommand\apjl{\ref@jnl{ApJ}}
\newcommand\apjs{\ref@jnl{ApJS}}
\newcommand\ao{\ref@jnl{Appl.~Opt.}}
\newcommand\apss{\ref@jnl{Ap\&SS}}
\newcommand\aap{\ref@jnl{A\&A}}
\newcommand\aapr{\ref@jnl{A\&A~Rev.}}
\newcommand\aaps{\ref@jnl{A\&AS}}
\newcommand\azh{\ref@jnl{AZh}}
\newcommand\baas{\ref@jnl{BAAS}}
\newcommand\jrasc{\ref@jnl{JRASC}}
\newcommand\memras{\ref@jnl{MmRAS}}
\newcommand\mnras{\ref@jnl{MNRAS}}
\newcommand\pra{\ref@jnl{Phys.~Rev.~A}}
\newcommand\prb{\ref@jnl{Phys.~Rev.~B}}
\newcommand\prc{\ref@jnl{Phys.~Rev.~C}}
\newcommand\prd{\ref@jnl{Phys.~Rev.~D}}
\newcommand\pre{\ref@jnl{Phys.~Rev.~E}}
\newcommand\prl{\ref@jnl{Phys.~Rev.~Lett.}}
\newcommand\pasp{\ref@jnl{PASP}}
\newcommand\pasj{\ref@jnl{PASJ}}
\newcommand\qjras{\ref@jnl{QJRAS}}
\newcommand\skytel{\ref@jnl{S\&T}}
\newcommand\solphys{\ref@jnl{Sol.~Phys.}}
\newcommand\sovast{\ref@jnl{Soviet~Ast.}}
\newcommand\ssr{\ref@jnl{Space~Sci.~Rev.}}
\newcommand\zap{\ref@jnl{ZAp}}
\newcommand\nat{\ref@jnl{Nature}}
\newcommand\iaucirc{\ref@jnl{IAU~Circ.}}
\newcommand\aplett{\ref@jnl{Astrophys.~Lett.}}
\newcommand\apspr{\ref@jnl{Astrophys.~Space~Phys.~Res.}}
\newcommand\bain{\ref@jnl{Bull.~Astron.~Inst.~Netherlands}}
\newcommand\fcp{\ref@jnl{Fund.~Cosmic~Phys.}}
\newcommand\gca{\ref@jnl{Geochim.~Cosmochim.~Acta}}
\newcommand\grl{\ref@jnl{Geophys.~Res.~Lett.}}
\newcommand\jcp{\ref@jnl{J.~Chem.~Phys.}}
\newcommand\jgr{\ref@jnl{J.~Geophys.~Res.}}
\newcommand\jqsrt{\ref@jnl{J.~Quant.~Spec.~Radiat.~Transf.}}
\newcommand\memsai{\ref@jnl{Mem.~Soc.~Astron.~Italiana}}
\newcommand\nphysa{\ref@jnl{Nucl.~Phys.~A}}
\newcommand\physrep{\ref@jnl{Phys.~Rep.}}
\newcommand\physscr{\ref@jnl{Phys.~Scr}}
\newcommand\planss{\ref@jnl{Planet.~Space~Sci.}}
\newcommand\procspie{\ref@jnl{Proc.~SPIE}}

\makeatother

\title{Simulation of diffusive particle propagation and related TeV $\gamma$-ray emission at the Galactic Center}
\author{\speaker{Alexander Ziegler} and Christopher van Eldik\\
Erlangen Centre for Astroparticle Physics (ECAP), Universit\"at Erlangen-N\"urnberg, Germany\\
E-mail: \email{alexander.ziegler@fau.de} 
}
\FullConference{The 34th International Cosmic Ray Conference,\\
	30 July- 6 August, 2015\\
	The Hague, The Netherlands
}
\ShortTitle{Simulation of diffuse TeV $\gamma$-ray emission at the Galactic Center}

\abstract{
Observations of the Galactic Center (GC) region in very-high-energy (VHE, $>\unit[100]{GeV}$) $\gamma$-rays, conducted with the High Energy Stereoscopic System (H.E.S.S.), led to the detection of an extended region of diffuse $\gamma$-ray emission in 2006.
To date, the exact origin of this emission has remained unclear, 
although a tight spatial correlation between the density distribution of the molecular material of the Central Molecular Zone (CMZ) and the morphology of the observed $\gamma$-ray excess points towards a hadronic production scenario.\newline
In this proceeding, we present a numerical study of the propagation of high-energy cosmic rays (CRs) through a turbulent environment such as the GC region. 
In our analysis, we derive an energy-dependent parametrization for the diffusion coefficient which we use for our simulation of the diffuse $\gamma$-ray emission at the GC. 
Assuming that hadronic CRs have been released by a single impulsive event at the center of our Galaxy, we probe the question whether or not the interaction processes of the diffusing hadrons with ambient matter can explain the observed diffuse $\gamma$-ray excess.
Our results disfavor this scenario, as our analysis indicates that the diffusion process is, on timescales compared to the typical proton lifetime at the GC region, too slow to explain the extension of the observed emission.   
}

\begin{document}
	
\section{Introduction}	
The GC region, harboring the supermassive black hole Sagittarius A* (Sgr A*) at its center, surrounded by massive clouds of dense molecular material, provides a unique opportunity for the study of non-thermal processes like the acceleration and the transport of highly energetic particles in our direct proximity.
Besides the detection of the two point-sources HESS~J1745-290 and G0.9+0.1 \cite{Aharonian2004,Aharonian2005}, ongoing observations of this region with the H.E.S.S.\ instrument led to the discovery of an extended region of diffuse TeV $\gamma$-ray emission, spanning a range of \textasciitilde$2^{\circ}$ in Galactic longitude.~A spatial correlation between the density distribution of the molecular material of the CMZ and the morphology of this diffuse emission points to a hadronic production scenario: locally accelerated hadronic CRs diffusing out of the GC might create the observed diffuse $\gamma$-ray flux through interaction processes with the material in the clouds leading to subsequent $\pi^{0}\rightarrow \gamma\gamma$ decays.

In the discovery paper \cite{2006Nature}, the H.E.S.S.\ collaboration argued that, assuming a mean diffusion coefficient of \textasciitilde $\unit[10^{30}]{cm^{2}/s}$, a single supernova explosion around $10^4$ years ago could have delivered enough energy to produce such a local population of highly energetic CRs, diffusing fast enough to produce a widely extended emission region comparable to the observed one.
Further studies followed up on this idea, deriving diffusion coefficients by fitting the output of high-energy CR propagation models to the characteristics of the observed emission \cite{Dimi2009,Busching2007}, with results in agreement with the H.E.S.S. proposal.
A fundamental different approach to the problem is given in \cite{Wommer2008}. Tracking individual particles under influence of the Lorentz force in turbulent magnetic fields, the outcome of this study suggests that particle diffusion is too slow to produce an emission region with an extension of more than a fraction of a degree at the GC, implying that diffusion coefficients should be much smaller than derived in previous studies. Hence, the conclusion of these authors is that only stochastic CR acceleration by magnetic turbulence throughout the GC region could produce the diffuse $\gamma$-ray excess - a scenario pursued in more detail e.g. in \cite{Fatuzzo2012}. 
Yet another approach assumes that a powerful wind, maintained by regular supernova explosions at the GC, advects particles out of the inner region, leading to a superposition of the diffusion-driven propagation with a ballistic component \cite{Crocker2011}.

A general overview of the most recent $\gamma$-ray observations of the GC region and related topics, including a detailed discussion of the detected diffuse TeV $\gamma$-ray emission can be found in \cite{vanEldik}. 
In summary, there is still no agreement on the question of how highly energetic particles get injected in the GC region and which of the above discussed mechanisms explains best the diffuse radiation.
In this proceeding, we focus on the scenario of a single impulsive injection of hadronic CRs at the center of our Galaxy, pursuing a preferably consistent treatment of the problem. From the tracking of ensembles of particles in turbulent magnetic fields, we derive a parametrization of the energy-dependence of the diffusion coefficient. 
The derived result is used as input parameter for our simulation of the diffuse $\gamma$-ray emission at the GC.
Discretizing the diffusion equation with the help of finite differences, our simulation works on a discrete three-dimensional spatial grid, tracking proton distributions of defined energy in discrete time steps. Embedded in this environment is a three-dimensional map of the molecular material of the CMZ, which allows us to take interaction processes into account, 
leading to TeV emission that we compare to the H.E.S.S.\ observations.

\section{Numerical analysis of charged-particle motion in turbulent magnetic fields} 
The aim of our numerical analysis presented here was to derive diffusion coefficients adjusted to the environmental conditions at the GC region, providing a statistical description of the scenario discussed above.
To this end, we follow the trajectories of individual particles solving the Lorentz force equation, assuming the propagation of test particles in an external magnetic field $\bm{B}$\,:
\begin{equation}\label{eq:lorentz}
\frac{\text{d}}{\text{d}t}\, \bm{p} = q \left(\bm{v}\times\bm{B}\right)\ ,
\end{equation} 
with the momentum $\bm{p}$, the velocity $\bm{v}$ and the charge $q$ of the considered particle. The time is denoted by $t$. For our numerical analysis, we take protons as representatives for the diffusing hadrons, since protons are by far the most abundant species of the known, local CRs.

\subsection{The turbulent magnetic field}
To numerically model the turbulent magnetic field, we adopt the method introduced by \cite{Giacalone1994}, which has already been applied in several studies on high-energy CR diffusion (e.g.\ \cite{Wommer2008,Fatuzzo2012,Fatuzzo2010}).   
Following \cite{Giacalone1994}, the total magnetic field $\bm{B}$ can be written as the sum of two components: a homogeneous background component $\bm{B}_0$, superimposed by a spatially fluctuating component $\delta\bm{B}(\bm{r})$. 
The turbulent component $\delta\bm{B}(\bm{r})$ is modeled as the sum of a large number $N$ of randomly polarized transverse waves 
\begin{equation}
\delta\bm{B}(\bm{r}) = \sum_{n=1}^N B_n \left[ \cos \alpha_n\, \hat{x}^{'} \pm i\, \sin \alpha_n\, \hat{y}^{'} \right] \times \text{exp}\left[ik_nz^{'}+i\beta_n\right]  
\end{equation}
where $\alpha_n$ and $\beta_n$ are random numbers between $0$ and $2\pi$. The values of the wave numbers $k_n$ are chosen evenly spaced on a logarithmic scale between $k_1\,=\,2\pi/ \lambda_\text{max}$ and $k_N\,=\,2\pi/ \lambda_\text{min}$, with $\lambda_\text{max}$ and $\lambda_\text{min}$ the maximum and minimum turbulent wavelength of the turbulent field. A random choice of the sign $\pm$ determines the helicity of the wavevector $\bm{k}_n$ about the $z^{'}$ axis. The laboratory frame (unprimed system) is related to the local coordinate system of an individual wave (primed system) via a rotation matrix $\bm{R}(\phi_n,\theta_n)$, where the propagation direction of each wave is chosen randomly: $0\leq\phi_n\leq2\pi$ and $0\leq\theta_n\leq\pi$.  
Hence, there are in total 5 random numbers needed for each value of $n$.
The spectrum of the turbulent component is defined by the applied turbulence model:
\begin{equation}
B_n^2 = B_1^2\left[\frac{k_n}{k_1}\right]^{-\Gamma+1} \ .
\end{equation} 
The results presented here were derived for a Kolmogorov-type spectrum, i.e.\ $\Gamma=5/3$. 
The overall normalization of the amplitudes is given by the relative energy density contained in the turbulent magnetic field with respect to the homogeneous background field.
For the case of a purely turbulent magnetic field (only taking $\delta\bm{B}$ into account), we define its total energy density to be equal to that of a homogeneous magnetic field of a certain desired magnetic field strength ${B}_0$. 

\subsection{Simulation procedure and definition of output measures}
For our simulation, we define particle ensembles of $N_p=1000$ protons of fixed energy $E$, which we track through the magnetic field configuration.
As primary output measures, we use the RMS $\sigma$ of the spatial distribution of the particles along the three spatial axes.
The value of the diffusion coefficient is extracted by fitting the time development of these quantities according to the expected relation $\sigma=\sqrt{2Dt}$, with $D(E)$ being the diffusion coefficient at the considered proton energy.
From Fig.\ \ref{FitD} it is visible that the particles need a certain period of time $t$\,\textasciitilde$\,\lambda_\text{max}/c$ to fully sample the turbulent structure of the field, until they reach the diffusive regime. From this point on, the RMS shows the expected time development $\sigma \propto \sqrt{t}$. 
We note that the observation of this transition at the observed point in time is consistent with various other studies on CR diffusion (e.g.\ \cite{Fatuzzo2012,Fatuzzo2010}).
For each energy value, 50 such simulations are performed.
This results in a set of 50 estimated values for $D$ of which we take the mean value and average over the three spatial directions.

As noted by \cite{Fatuzzo2010}, only the value chosen for the maximum wavelength $\lambda_\text{max}$ with respect to the particle gyroradius $R_g$ has an significant impact on the output measures.
Therefore, for the results presented here, the remaining parameters were set to the following specifications: $N=200$ values of $k$ to set up the turbulent field and a minimum turbulent wavelength of $\lambda_\text{min} = 0.1 R_g$. These settings are similar to those presented in \cite{Wommer2008,Fatuzzo2012}. 
To adjust the formalism to the conditions of the GC region, we use a magnetic field strength of $\unit[50]{\mu G}$, in agreement with the lower limit estimate for the GC magnetic field strength given by \cite{Crocker2010}.
As there is no conclusive picture on the exact magnetic field configuration, we restrict the results presented here to the case of a purely turbulent magnetic field.
As discussed in \cite{Fatuzzo2010}, the value of $\lambda_\text{max}$ should be constrained from above by the entire spatial extension of the molecular clouds (typically a few tens of parsecs), and from below by the size of their dense cores (only fractions of a parsec). 
In accordance with the studies \cite{Fatuzzo2012,Fatuzzo2010}, we chose an intermediate value of $\lambda_\text{max}=\unit[1]{pc}$.

\subsection{Derived diffusion coefficients}
\begin{figure}[t!]
	\begin{minipage}[t]{0.475\textwidth}
		\centering
		\includegraphics[width=80mm]{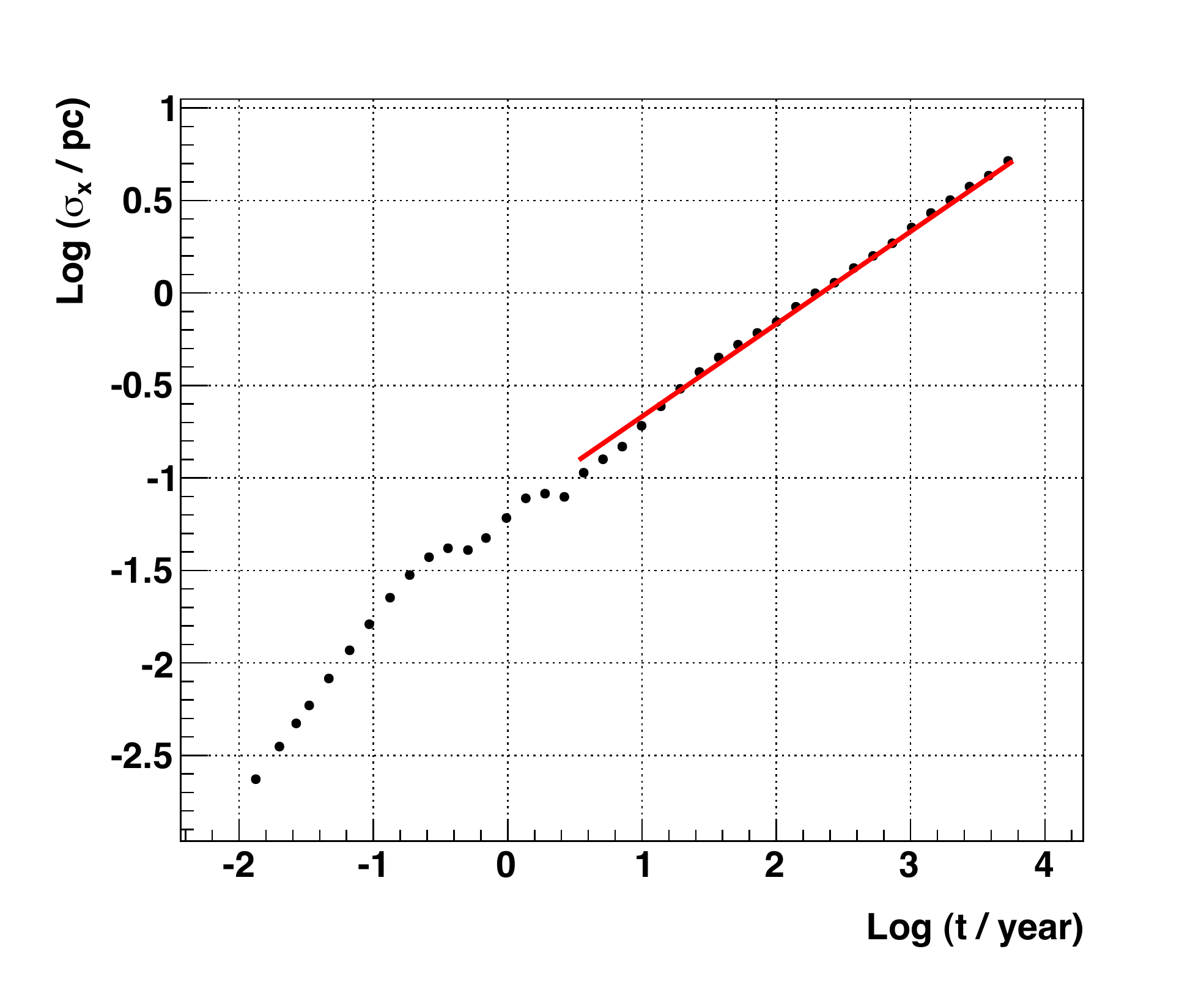}			 		
		\caption{Time development of the output measure ${\sigma}_{x}$ for a proton energy of $E=\unit[3\times10^{15}]{eV}$. The red line shows the fit to the data according to the expected time evolution $\sigma=\sqrt{2Dt}$.
			The lower fit boundary is set to $t=\lambda_\text{max}/c$  
			(\textasciitilde\,0.5 on the shown log-scale for $\lambda_\text{max} = \unit[1]{pc}$).
		} 
		\label{FitD}
	\end{minipage}
	\hspace{0.5cm}
	\begin{minipage}[t]{0.45\textwidth}
		\centering
		\includegraphics[width=80mm]{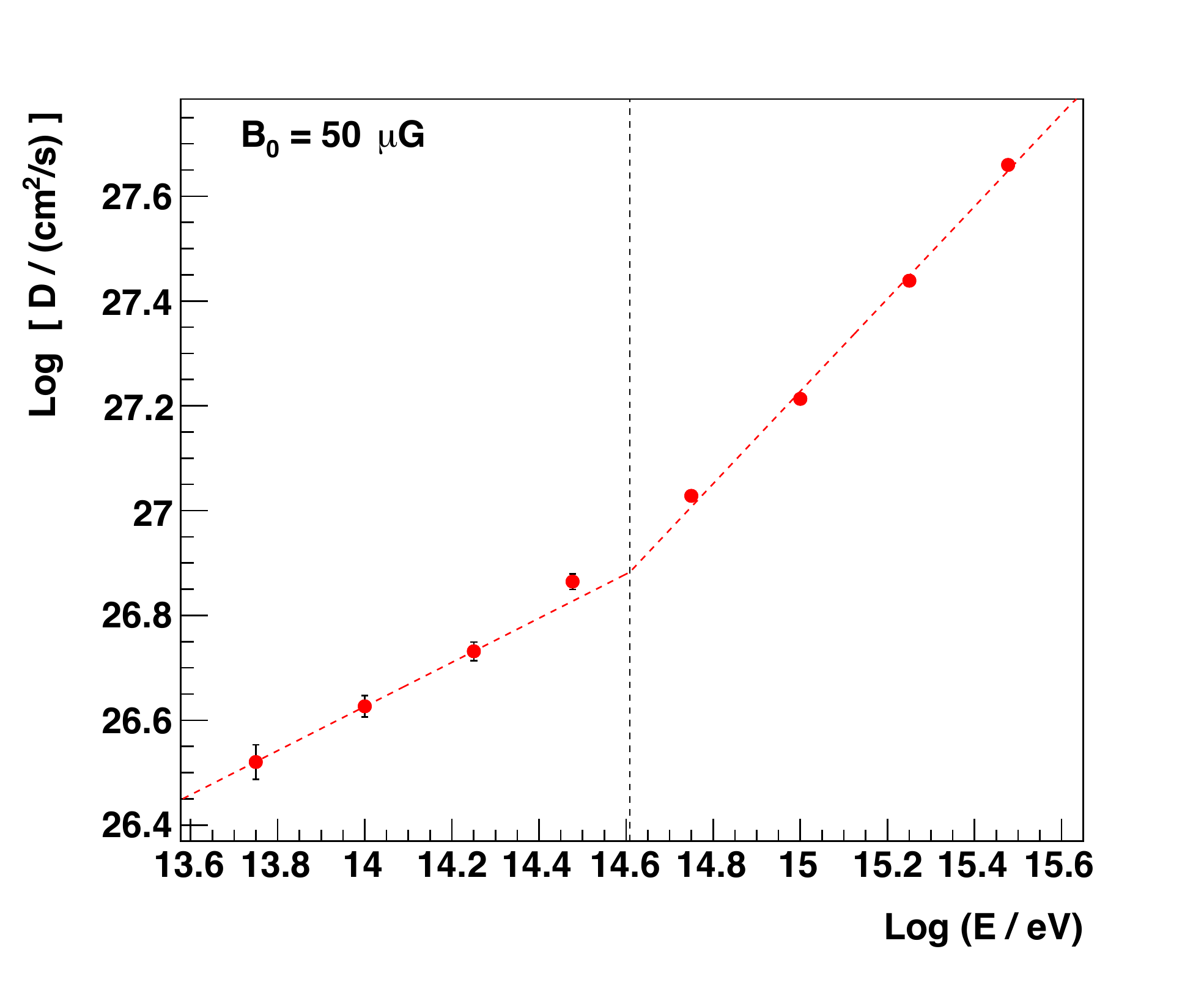}
		\caption{Diffusion coefficient as a function of energy, derived for a purely turbulent magnetic field with a Kolmogorov-type power spectrum. A clear break in the energy-dependence is visible, indicating two different dynamical regimes for the explored energy range.}
		\label{diffcoeffs}
	\end{minipage}
\end{figure}

We present here the diffusion coefficients we derived assuming a purely turbulent magnetic field, generated according to the specifications above. The results are shown in Fig.\ \ref{diffcoeffs}. It is usually assumed that the diffusion coefficient scales as a function of energy with a power law with a constant spectral index, see e.g.\ \cite{Aharonian1996,Gabici2009}.
The results presented here show, however, a clear break in the energy dependence of $D(E)$ in the explored energy range. This result is consistent with the studies presented in \cite{Fatuzzo2010} for a Kolmogorov-type spectrum.
We assume that we can model both regimes (above and below the break) by a power law with a constant spectral index, separately. 
Doing so, we find the position of the break at $\epsilon$\,\textasciitilde\,$0.009$, following \cite{Fatuzzo2010} who introduced the dimensionless parameter $\epsilon=E/E_{0}$ with $E_{0}$ the energy of a proton with a gyroradius equal to $\lambda_\text{max}$ ($E_{0}=4.6\times10^{16}\,$eV for the chosen settings).
This is close to the break position quoted by \cite{Fatuzzo2010}, $\epsilon$\,\textasciitilde\,$0.005$, who argue that the spectral break is related to a change in the dynamics of the particles:
for $\epsilon\ll\epsilon_\text{break}$, they are tightly coupled to the magnetic field lines and their diffusion is strongly dictated by the field structure, whereas with growing energy (for $\epsilon>\epsilon_\text{break}$), particles start to decouple from the field lines and diffuse on length scales defined by the value of their gyroradii.

The parameters we derive with linear fits to the data points on the shown double logarithmic scale are summarized in Tab.\ \ref{tab:fitres} (the data point at energy $\text{Log}(E/\unit[]{eV})\approx 14.5$ is excluded from the fit procedure to not depend too strongly on the exact shape close to the break).  
The value we derive for the diffusion coefficient at energy $10\,$GeV, $D_{10}$, is relatively small compared to the usually assumed range of $D_{10}\,$\textasciitilde$\,\unit[10^{26}-10^{28}]{cm^2/s}$, where smaller values refer to dense regions of interstellar gas and \textasciitilde$\unit[10^{28}]{cm^2/s}$ refers to the commonly proposed value for the interstellar medium (e.g.\ \cite{Aharonian1996,Gabici2009}).
The value for the spectral index in the low energy regime, $\delta = 0.42\pm0.07$, lies within the expected range $\delta$\,\textasciitilde$\ 0.3 - 0.6$ (see e.g.\ \cite{Aharonian1996,Gabici2009}). In the high energy regime we derive a higher value of $\delta~=~0.88\pm0.02$.
\begin{table}[h!]
	\begin{center}
		\caption{
			Obtained fit parameters for the energy dependence of the diffusion coefficient, 
			shown in Fig.\ \protect\ref{diffcoeffs}. See text for further details on the parameters chosen in the simulation.
			}
		\label{tab:fitres}
		\vspace{0.2cm}
		\begin{tabularx}{0.5\textwidth}{r|c|c}
			      &  Log[ $D_{10}$/($\text{cm}^2/\text{s}$) ] & $\delta$ \\
			\hline
			\multirow{2}{*}{ }
			 $\epsilon\leq\epsilon_\text{break}$			& $24.9\pm1.2$		& $0.42\pm0.07$			\\
			 $\epsilon>\epsilon_\text{break}$	    	    & $22.8\pm0.3$		& $0.88\pm0.02$			\\
		\end{tabularx}
	\end{center}
\end{table}

\section{Simulation of the diffuse TeV $\gamma$-ray emission at the Galactic Center}
For our separate simulation of the diffuse TeV $\gamma$-ray emission at the GC, we use the above derived energy-dependent diffusion coefficient as input parameter. We assume a single impulsive injection of protons at the center of our Galaxy at a certain point of time in the past and calculate today's $\gamma$-ray emission resulting from the interactions of the diffusing particles with the ambient matter. 

\subsection{The model}
The developed simulation is based on the assumption of spherical symmetric diffusion, combined with a discretization of the diffusion equation applying the finite-difference method.
It works on a discrete three-dimensional spatial grid, tracking proton distributions of defined energy in discrete time steps.
Within this environment we embedded a three-dimensional density map of the molecular material of the CMZ, based on the CS line emission measurements presented in \cite{Tsuboi1999}. For a line-of-sight positioning of the cloud centers, we use the results provided by \cite{Sawada2004}.
The spatial calculation grid comprises $\pm\,525\,$pc in each spatial direction, the proton source is located at the origin. The bin width is $10\,$pc (\textasciitilde$\,0.07^\circ$) in each direction.
The energy range we take into account is $10^{11}-10^{17}\,\text{eV}$,  the differential proton energy spectrum is assumed to follow a power law with constant spectral index: ${\text{d}N_p/\text{d}E_p\propto E_p^{-2}}$. 
The applied formalism provides a very flexible treatment of the addressed scenario, given the information of local proton populations for each spatial bin at each point in time.

\begin{figure}[t!]
	\begin{minipage}[t]{0.475\textwidth}
		\centering
		\includegraphics[width=75mm]{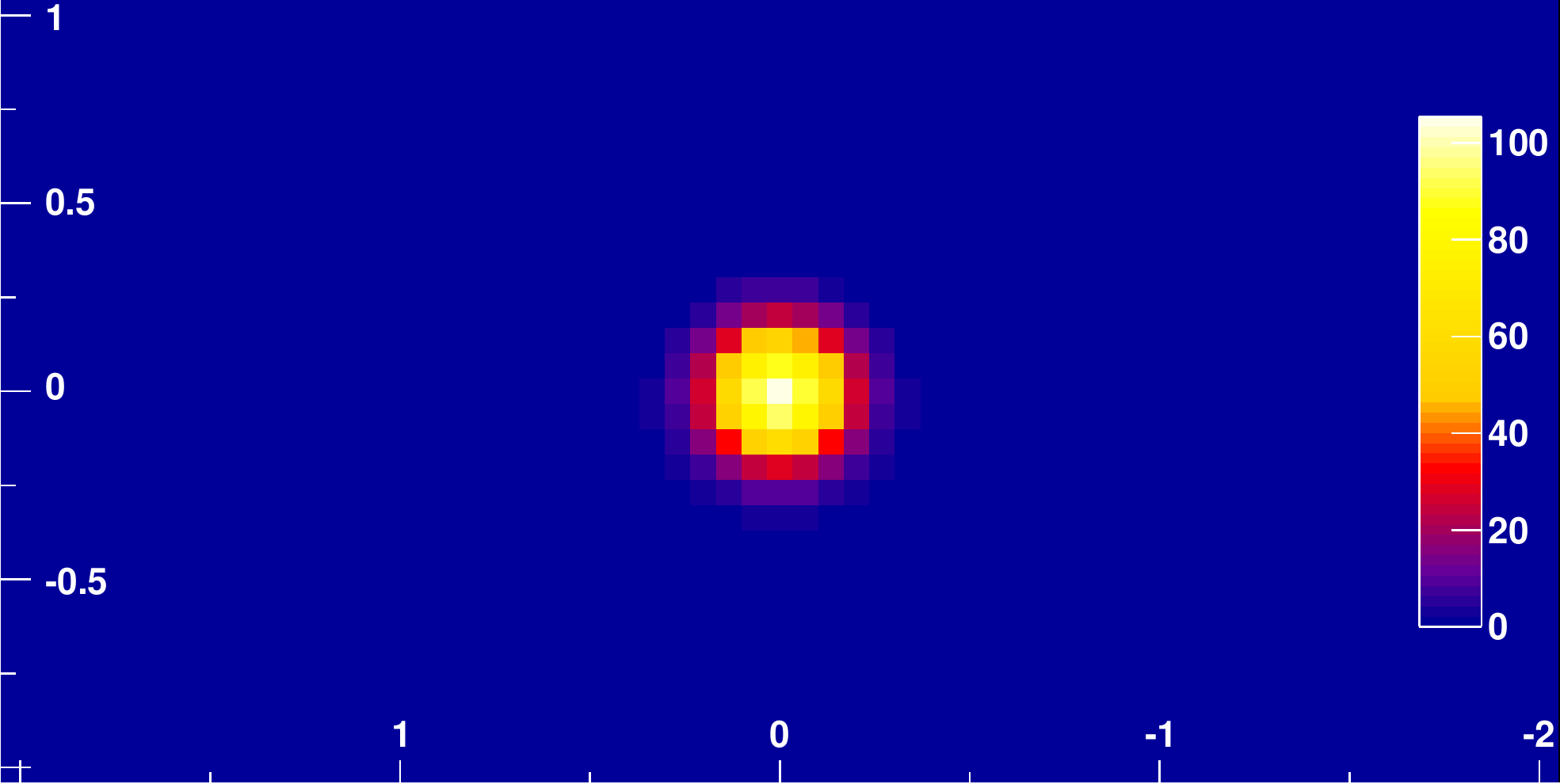}			 		
	\end{minipage}
	\hspace{0.5cm}
	\begin{minipage}[t]{0.45\textwidth}
		\centering
		\includegraphics[width=75mm]{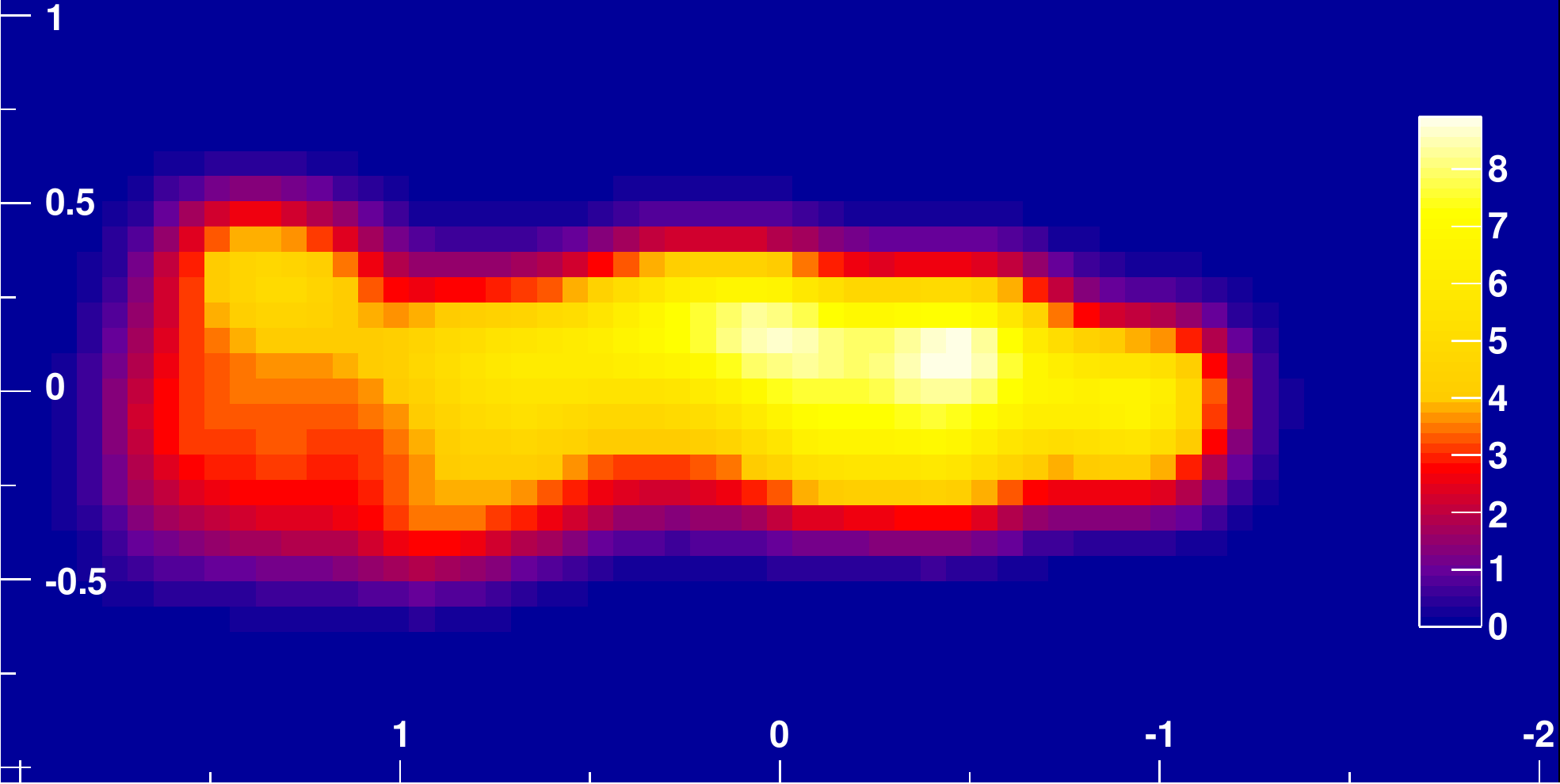}
	\end{minipage}
	\caption{$\gamma$-ray count maps of the GC region calculated with the simulation using the diffusion coefficient shown in Fig.\ \protect\ref{diffcoeffs}.
	Axes are Galactic longitude (x) and Galactic latitude (y).
	Left (right): total tracking time of $10^{4}$ ($10^{7}$) years.	
	For the longer propagation time, the emission region broadens significantly. However, a very high source energy (\textasciitilde$\unit[10^{55}]{erg}$) would be required to explain the measured $\gamma$-ray flux spectrum, making the proposed scenario nevertheless rather unlikely.}
	\label{countmaps}
\end{figure}

\subsection{Results}   
The resulting $\gamma$-ray count map we derive for a total tracking time of $10^{4}$ years (the diffusion timescale proposed in \cite{2006Nature}) is shown in Fig.~\ref{countmaps}, left. Obviously, the large spatial extension of the measured $\gamma$-ray excess, about $2^\circ$ Galactic longitude $\times$ $0.5^\circ$ Galactic latitude~(cf.~\cite{2006Nature}), cannot be reproduced, as the protons diffuse away too slowly from their point of origin.
Instead, the morphology of the derived emission region is comparable to the emission characteristic of a point source, with a diameter (full width at half maximum) of \textasciitilde\,$0.2^\circ$.
In order to provide the opportunity for large-scale $\gamma$-ray emission, we extended our analysis to larger tracking timescales. Our results indicate that for propagation periods of \textasciitilde$10^7$ years (Fig.\ \ref{countmaps}, right), the emission region reaches a spatial scale comparable to the measured one.
 
We estimate the total required source energy $\omega_\text{tot}$ by fitting the calculated $\gamma$-ray flux to the flux measured with the H.E.S.S. instrument \cite{2006Nature}.
The values discussed in the following refer to the energy interval of $10^{9}-10^{15}\,\text{eV}$.
For this interval, the H.E.S.S. collaboration estimated for the discussed scenario a value of \textasciitilde$\unit[10^{50}]{erg}$ required to accelerate the local proton population, which translates to a total source energy of $\omega_\text{tot}$\textasciitilde$\unit[10^{51}]{erg}$ assuming $10\%$ efficiency for particle acceleration.
This value is in good agreement with the scenario of a supernova explosion, which yields a typical energy release of \textasciitilde$\unit[10^{51}]{erg}$.
For a tracking time of $10^{4}$ years, we calculate a value of \textasciitilde$\unit[10^{49}]{erg}$ required by the proton spectrum, which translates to an acceleration efficiency of only 1\% assuming a total source energy of \textasciitilde$\unit[10^{51}]{erg}$.
For the larger timescales, we derive however significantly larger source energies, due to a growing fraction of particle losses during propagation. The proton lifetime can approximately be written as ${\tau_{pp}\approx3\times10^{5}\,\text{yr}\,\left(n_{H_2}/(100\,\text{cm}^{-3})\right)^{-1}}$~(e.g.\ \cite{Fatuzzo2010}). It follows that for timescales much larger than $10^{5}$~years, a significant fraction of the initial particle population will be lost until the interested point in time due to interaction processes.
Therefore, the total source energy we calculate for a timescale of $10^7$ years is, with a value of \textasciitilde$\unit[10^{55}]{erg}$ (assuming 10\% acceleration efficiency), significantly larger than the typical explosion energy delivered by a supernova explosion.  
The high energy value we derive here makes the discussed scenario rather unlikely, even though the morphology we obtain for the $\gamma$-ray excess for this timescale might match the essential characteristics of the measured one.

\section{Conclusion}
In this proceeding, we have presented our studies on the scenario of a production of the diffuse VHE $\gamma$-ray emission at the GC via interaction processes of diffusing protons with the ambient molecular matter of the CMZ.
Within this scenario, we have assumed that a local population of highly energetic CRs had been released by a single impulsive ejection of a single power source located at the GC.\newline

To this end, we have numerically analyzed the motion of protons in turbulent magnetic fields, applying the formalism introduced in \cite{Giacalone1994}, adjusted to the environmental conditions at the GC region. Tracking individual particles in a purely turbulent magnetic field by solving the Lorentz force equation, we derive an energy-dependent parametrization for the diffusion coefficient. In agreement with the analysis presented in \cite{Fatuzzo2010}, we observe two different dynamical regimes in the explored energy range. 
Modeling both regimes independently according to the usual assumption $D(E)\propto E^\delta$, we derive a value for $D_{10}$ below (but in agreement with) the typically assumed range of $D_{10}\,$\textasciitilde$\ \unit[10^{26}-10^{28}]{cm^2/s}$. 
The values we derive for the spectral index are $\delta=0.42$ in the low energy range, and $\delta=0.88$ in the high energy regime.
We use this parametrization of $D(E)$ to simulate the diffuse $\gamma$-ray emission according to the described scenario.  
For a total propagation time of $10^4$~years, the excess region exhibits a spatial extension much smaller than the observed one, due to too slow diffusion of the underlying particle distribution.
For significantly larger timescales, particle losses due to interaction processes during propagation lead to very high estimates of the total source energy (\textasciitilde$\unit[10^{55}]{erg}$), required to match the measured flux of the diffuse $\gamma$-ray excess.\newline

In summary, we have found that diffusion of hadronic CRs away from a single point source at the GC appears to be too slow to explain the diffuse $\gamma$-ray excess measured by H.E.S.S., extending about $2^\circ$ in Galactic longitude.
This result is in agreement with the analysis presented in \cite{Wommer2008}.
Possibly, the $\gamma$-ray emission of the central source HESS J1745-290 itself might be associated with the addressed scenario.
Besides the requirement of further refined simulation studies, the planned Cherenkov Telescope Array (see e.g.\ \cite{CTA2011}) will help to improve our understanding of the question which of the possible remaining scenarios, including stochastic particle acceleration as well as particle advection by a galactic wind outblow, might be the correct one to explain the current VHE $\gamma$-ray observations of the GC and its related particle physics.

\end{document}